\documentclass[pra,twocolumn,showpacs,showkeys]{revtex4}
\usepackage{amsmath}
\usepackage{amsfonts}
\usepackage{graphicx}

\begin{document}

\title{Elementary modes of excitation caused by the quadratic Zeeman term
and the sensitivity of spin structures of small spin-2 condensates against
the magnetic field}
\author{Y. Z. He}
\author{C. G. Bao}
\thanks{Corresponding author: stsbcg@mail.sysu.edu.cn}
\affiliation{Center of Theoretical Nuclear Physics, National Laboratory of Heavy Ion
Accelerator, Lanzhou, 730000, People's Republic of China}
\affiliation{State Key Laboratory of Optoelectronic Materials and Technologies, School of
Physics and Engineering, Sun Yat-Sen University, Guangzhou, 510275, People's
Republic of China}
\date{\today}

\begin{abstract}
The response of spin-2 small condensates to an external magnetic field $B$
is studied. The parameters of the interaction are considered as variable.
The emphasis is placed on clarifying the modes of excitation caused by the
quadratic Zeeman term. The theoretical method used is beyond the mean field
theory. A set of eigenstates with the $U(5)\supset SO(5)\supset SO(3)$
symmetry is introduced to facilitate the analysis. To obtain a quantitative
evaluation on the response, the fidelity susceptibility and the $B$
-dependent average populations of spin-components have been calculated.
Mostly the particle number $N=30$ is assumed. The effect with a larger or
smaller $N$ is also considered. It was found that the sensitivity of the
response depends strongly both on the interaction and on the inherent
symmetry.
\end{abstract}

\pacs{03.75.Mn, 03.75.Hh}
\keywords{spin correlation, Bose-Einstein condensates, spin structures,
spin-dependent force of atoms}
\maketitle

\section{Introduction}

The physical properties of Bose-Einstein condensates of atoms
with nonzero spin can be tuned by external laser fields and
magnetic fields \cite{r_SDM1998}. These condensates are much
less affected by defects and impurities which exist extensively
in usual condensed matter systems. Therefore, the condensates
can be used to simulate a variety of ideal condensed matter
systems (e.g., the system of gluons) for academic studies, or
to work as ideal magnetic materials in practical application
(say, in quantum information and quantum computation). For the
purpose of manipulation, how the condensates will react to an
external magnetic field is a crucial problem to be clarified.

Usually, the particle number $N$ in a condensate would $\geq 10^4$. If $N$
could be greatly reduced, new physics might emerge. For an example, for
spin-2 condensates, it was predicted based on the mean field theory (MFT)
that the ground state would have three phases, namely, the ferromagnetic
phase ($f$-phase), polar phase ($p$-phase), and cyclic phase ($c$-phase)
\cite{r_CCV2000,r_KM2000,r_UM2002}. However, from a study based on a theory
beyond the MFT, the ground state may have 3, 4, or 5 phases depending on $N$
\cite{r_IPV2007}. The fine effect of $N$ could only be seen if $N$ is small
(say, $N\leq 100$). When $N$ is small, the properties of the condensates
might depend on $N$ sensitively, just as the properties of the nuclei depend
on the number of nucleons sensitively. On the other hand, recent progress in
technique suggests that the condensates with $N$ very small could be
experimentally achieved \cite{r_GN2010}. Therefore, a study of the small
condensates is worthy because new knowledge additional to those from large
condensates might be obtained.

This paper is dedicated to small spin-2 condensates, and is a
generalization of two previous papers \cite{r_HY2011,r_B2011}
on spin-2 and spin-1 condensates. The method adopted is beyond
the MFT. The aim is to study the sensitivity of the small
condensates responding to the variation of an external magnetic
field $B$. To this aim the fidelity susceptibility
\cite{r_QHT2006,r_YWL2007,r_ZP2007} and the average population
of spin-components have been calculated. To clarify the
underlying physics, the set of eigenstates with the
$U(5)\supset SO(5)\supset SO(3)$ symmetry firstly proposed in
the refs.~\cite{r_IPV2007,r_GA2004} is introduced. Based on the
set, elementary excitation modes caused by the quadratic Zeeman
term have been found. These modes together with the associated
energy gap are decisive to the response of the spin-structures
against the appearance of $B$. It is well known that the
structures of the ground states depend on the interaction. On
the other hand, the ways of excitation caused by the field is
found to be strongly constrained by the inherent symmetry. Thus
both the interaction and symmetry together determine the
response of the condensates to the field as shown below.

Since the study is not dedicated to a specific spin-2 system, the parameters
of the spin-dependent interaction are considered as variable. Mostly $N=30$
is given, the effect of a larger or smaller $N$ is also discussed.

\section{Hamiltonian}

The atoms are assumed to be confined by an isotropic and
parabolic potential $\frac{1}{2}m\omega^2 r^2$. The temperature
$T$ is assumed to be very low. Since the energy of spatial
excitation of individual atom is $\sim\hbar\omega$, when
$T<<\hbar\omega/k_B$ it is reasonable to assume that no atoms
would be spatially excited and all of them would fall into a
common spatial-state $\phi(\textbf{\textit{r}})$. This
assumption is called the single mode approximation (SMA), which
has been quite frequently used in the literatures. Furthermore,
a magnetic field $B$ lying along the $Z$-axis is set. When an
integration is carried out over the spatial degrees of freedom,
we arrive at a model Hamiltonian depending  only on the
spin-degrees of freedom as\cite{r_JS1998,r_IPV2007}
\begin{equation}
 H_{\mathrm{mod}}
  =  \sum_{i<j}
     V_{ij}
    -p
     \sum_i
     \hat{f}_{iZ}
    +q
     \sum_i
     (\hat{f}_{iZ})^2,
 \label{e01_Hmod}
\end{equation}
where an overall constant has been neglected, $V_{ij}=\sum_S
g_S\bar{n}\mathcal{P}_S^{ij}$, where $S$ is the combined spin
of the particles $i$ and $j$, $\mathcal{P}_S^{ij}$ is the
projector of the $S$-channel, $g_S$ is the strength of
interaction related directly to the $s$-wave scattering length
of the $S$-channel, $\bar{n}=\int|\phi(\textbf{\textit{r}})|^4
\mathrm{d}\textbf{\textit{r}}$, and the last two terms are for
the linear and quadratic Zeeman energies, respectively. Note
that $\mathcal{P}_4^{ij}
=1-\mathcal{P}_0^{ij}-\mathcal{P}_2^{ij}$. We further introduce
$\beta$ and $\theta$ so that $\beta\cos\theta=(g_0-g_4)/g_4$,
$\beta\sin\theta =(g_2-g_4)/g_4$. Then we define
$H'_{\mathrm{mod}} \equiv\frac{1}{\beta
g_4\bar{n}}H_{\mathrm{mod}}$. Furthermore, since $M$ (the
$Z$-component of the total spin) remains to be conserved under
the field $B$, the linear Zeeman term provides only a constant
and therefore can be neglected. When all the constants involved
are removed, $H'_{\mathrm{mod}}$ can be rewritten as
\begin{equation}
 H'_{\mathrm{mod}}
  =  \sum_{i<j}
     ( \cos\theta \
       \mathcal{P}_0^{ij}
      +\sin\theta \
       \mathcal{P}_2^{ij} )
    +q'
     \sum_i
     (\hat{f}_{iZ})^2,
 \label{e02_Hmodp}
\end{equation}
where $\theta$ is from 0 to $2\pi$, $q'=q/(\beta g_4\bar{n})$.
When the Hamiltonian is transformed from Eq.~(\ref{e01_Hmod})
to (\ref{e02_Hmodp}), the set of eigenenergies will be
multiplied by a common constant and will be further shifted as
a whole. But the eigenstates will remain unchanged. Since only
two parameters $\theta$ and $q'$ are contained in
$H_{\mathrm{mod}}'$, related analysis is easier to perform.

When $q'=0$, it has been proved that $H'_{\mathrm{mod}}$ can be
rewritten as a sum of four Casimir operators of a chain of
groups $U(5)\supset SO(5)\supset SO(3)$ \cite{r_IPV2007}.
Therefore the problem can be solved analytically based on the
group algebra \cite{r_IPV2007,r_GA2004,r_CE1976}. Three quantum
numbers, namely, $F$ (total spin), $M$ (its $Z$-component), and
$v$ (seniority, $N-v$ must be an even integer, and $(N-v)/2$ is
the number of singlet pairs) are introduced to classify the
states. The eigenenergies related to these quantum numbers are
\begin{equation}
 E_{v F}
  =  \frac{10\sin\theta-7\cos\theta}{70}
     v
     (v+3)
    -\frac{\sin\theta}{14}
     F
     (F+1),
 \label{e03_EvF}
\end{equation}
where an irrelevant overall constant depending on $N$ has been further
neglected. The associated eigenstates are denoted as $\Psi_{v FM}$. Note
that, for specifying a symmetry-adapted eigenstate, not all the $(v,F)$
pairs are allowed. Some of them are prohibited by inherent symmetry. The
details are referred to \cite{r_GA2004}. Eq.~(\ref{e03_EvF}) leads to a
notable feature, namely, when $\sin\theta =\frac{7}{10}\cos\theta$, all the
symmetry-adapted $\Psi_{v FM}$ with the same $F$ are degenerate. Whereas
when $\sin\theta=0$, all the symmetry-adapted $\Psi_{v FM}$ with the same $v$
are degenerate. If the group of degenerate states happens to be the ground
states, the low-temperature behavior of the condensate would be seriously
affected.

In general, a slight increase of $N$ (say from being even to
odd) will affect the structure of the ground state . However,
when $N$ is large, the effect is weak. For the convenience of
discussion, it is assumed that $N$ is a multiple of 2 and 3 in
the follows (other cases of $N$ can be similarly discussed).
When $\theta$ and $M$ are fixed ($M\geq 0$ is assumed), from
Eq.~(\ref{e03_EvF}) and from the symmetry-adaptability, the
pair $(v_g,F_g)$ for the ground state (the lowest state with
the given $M$) can be determined. When $M=0$, based on the
method of group algebra, there are three regions of $\theta$
\cite{r_IPV2007}. When $\theta$ is from $0$ to $\theta_{fp}
\equiv\arctan[-\frac{7(N+3)}{10(N-2)}]<\pi$ (region I), $v_g=N$
and $F_g=2N$, and accordingly the ground state is in the
$f$-phase. In this phase all the spins are nearly aligned along
a common direction to form a total spin, which is vertical to
the $Z$-axis (if $M=0$) and has an arbitrary azimuthal angle.
When $\theta$ is from $\theta_{fp}$ to $\theta_{pc}
\equiv\arctan(\frac{7}{10})=214.99^{\circ}$ (region II),
$v_g=0$ and $F_g=0$, and accordingly in the $p$-phase, in which
the spins are two-by-two to form the singlet pairs (If $N$ is
odd, the ground state would have $v_g=1$ and $F_g=2$, and
accordingly would be composed of a group of singlet pairs
together with a single particle. This case is not discussed in
this paper ). When $\theta$ is from $\theta_{pc}$ to $2\pi$
(region III), $v_g=N$ and $F_g=0$, and accordingly in the
$c$-phase, in which the spins are essentially three-by-three to
form the triplexes (If $N$ is not a multiple of 3, additional
particle(s) would be added)\cite{r_IPV2007}. The division into
three regions based on the group algebra coincides with that
from the MFT when $N$ is sufficiently
large\cite{r_CCV2000,r_UM2002}. Nonetheless, the details of
spin correlations can be understood from the former but can not
from the MFT.

When $q'\neq 0$, there is no analytical solution. $F$ is no
more conserved, but $M$ remains to be a good quantum number. To
obtain numerical solution, the Fock-states $|\alpha\rangle
=|N_2^{\alpha},N_1^{\alpha},N_0^{\alpha},N_{-1}^{\alpha},N_{-2}^{\alpha}\rangle$
are used as basis functions for the diagonalization of
$H'_{\mathrm{mod}}$, where $N_{\mu}^{\alpha}$ is the number of
particles in the $\mu$-spin-component,
$\sum_{\mu}N_{\mu}^{\alpha}=N$ and $\sum_{\mu}\mu
N_{\mu}^{\alpha}=M$, $|\alpha\rangle$ as a whole form a
complete set for symmetrized spin-states. Thereby exact
eigenstates $\psi_{iM}^{q'}$ of $H'_{\mathrm{mod}}$ can be
obtained, where $i$ is a serial number of the series of states
with the given $M$ ($i=1$ denotes the ground state). Various
information will be extracted from $\psi_{iM}^{q'}$. The case
with $M=0$ is firstly considered.

\section{The fidelity susceptibility of the ground state and the mode of
excitation}

\begin{figure}[htbp]
 \centering
 \resizebox{0.95\columnwidth}{!}{\includegraphics{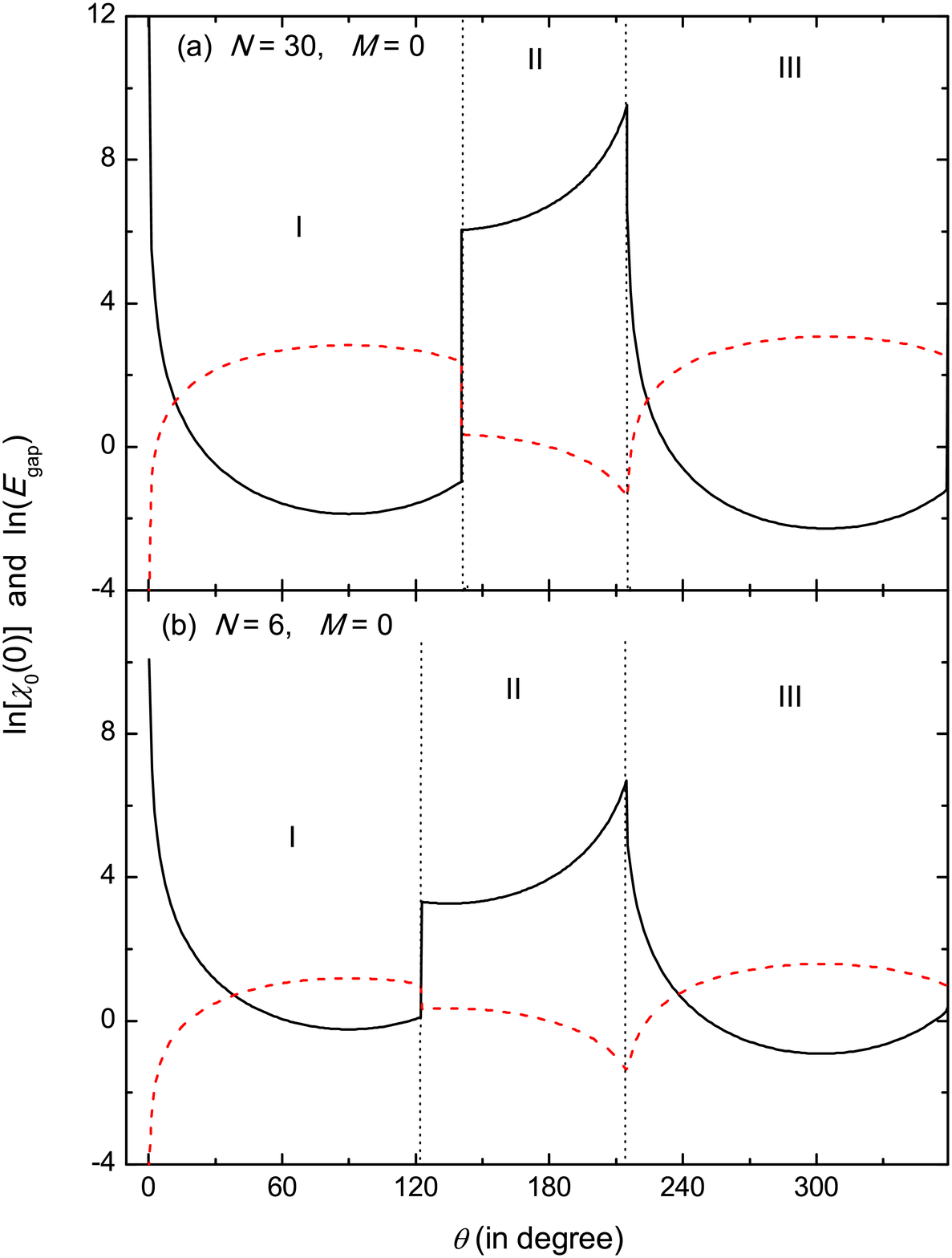}}
 \caption{(Color online.) $\ln[\chi_0(0)]$ (solid) and
$\ln(E_{\mathrm{gap}})$ (dash) against $\theta$. $M=0$ is
assumed. The two vertical dotted lines located at $\theta_{fp}$
and $\theta_{pc}$, respectively, are introduced\ for separating
the regions I, II, and III. In (a) $N=30$ and
$\theta_{fp}=140.48^{\circ}$, while in (b) $N=6$ and
$\theta_{fp}=122.41^{\circ}$. In both (a) and (b),
$\theta_{pc}=214.99^{\circ}$.}
 \label{fig1}
\end{figure}

Firstly, we would like to study the sensitivity of the ground
state $\psi_{1M}^{q'}$ against the variation of the field. For
this purpose the fidelity susceptibility
\cite{r_QHT2006,r_YWL2007,r_ZP2007}
\begin{equation}
 \chi_M(q')
  =  \lim_{\varepsilon\rightarrow 0}
     \frac{2}{\varepsilon^2}
     ( 1
      -|\langle
        \psi_{1M}^{q'+\varepsilon}|
        \psi_{1M}^{q'}
        \rangle| ),
 \label{e04_chiMqp}
\end{equation}
has been calculated. A larger $\chi_M(q')$ implies that the
deviation between $\psi_{1M}^{q'+\varepsilon}$ and
$\psi_{1M}^{q'}$ is larger, thus the state responds more
sensitively to $q'$. Therefore this quantity measures
quantitatively the sensitivity. From the eigenstates
$\psi_{1,0}^{\varepsilon}$ and $\psi_{1,0}^0$ (both can be
obtained via the diagonalization), $\ln[\chi_0(0)]$ of the
ground state as a function of $\theta$ is shown by the solid
curves in Fig.~\ref{fig1}. The behaviors of $\chi_0(0)$ in the
three regions of $\theta$, which are separated by the dotted
vertical lines, are different. In general, the $p$-phase in
region II is much more sensitive to the appearance of the
magnetic field than the $f$- and $c$-phases. From
Eq.~(\ref{e03_EvF}), we know that the ground state is
degenerate when $\theta=\theta_{pc}$ or 0. In these cases, the
susceptibility can not be well defined. Therefore, in
Fig.~\ref{fig1}, the two narrow domains
$(\theta_{pc}-\delta,\theta_{pc}+\delta)$ and
$(-\delta,\delta)$ with $\delta=\pi/180$ are actually not
included in the calculation. Nonetheless, when $\theta$ is
close to these narrow domains, the sensitivity is very high.

\begin{table*}[htbp]
\caption{The three phases of the ground states with $M=0$ and
the associated three connected regions of $\theta$. $N$ is
assumed to be a multiple of 2 and 3. $(v_g,F_g)$ are the
quantum numbers of the ground state, $(\bar{v}_g,\bar{F}_g)$
specifies the corresponding unique mode of elementary
excitation, and $E_{\mathrm{gap}}$ is the associated energy of
excitation. $Q_{\bar{v}_g\bar{F}_g,v_g F_g}^0$ is the matrix
element of the elementary excitation caused by the quadratic
Zeeman term $\sum_i(\hat{f}_{iZ})^2$. This matrix element is
given at $N=30$ and 6 (inside the parentheses).}
\begin{ruledtabular}
  \label{tab1}
  \begin{tabular}{llll}
  region                                    & I (ferromagnetic)
  & II (polar)                              & III (cyclic) \\
  \hline
  right border                              & $\theta_{fp}=\arctan[\frac{-7(N+3)}{10(N-2)}]<\pi$
  & $\theta_{pc}=\arctan(\frac{7}{10})>\pi$ & $\theta_{cf}=2\pi$ \\
  $v_g,F_g$                                 & $N,2N$
  & $0,0$                                   & $N,0$ \\
  $\bar{v}_g,\bar{F}_g$                     & $N,2N-2$
  & $2,2$                                   & $N-2,2$ \\
  $E_{\mathrm{gap}}$                        & $\frac{4N-1}{7}\sin\theta$
  & $\sin\theta-\cos\theta$                 & $\frac{7\cos\theta-10\sin\theta}{70}(4N+2)-\frac{3\sin\theta}{7}$ \\
  $Q_{\bar{v}_g\bar{F}_g,v_g F_g}^0$        & 6.68 (2.93)
  & 28.98 (7.27)                            & 6.93 (3.10)
 \end{tabular}
 \end{ruledtabular}
\end{table*}

In order to understand the behavior shown in Fig.~\ref{fig1}, we have to
study the mode of excitation caused by the quadratic Zeeman term. The rule
of selection governing the matrix elements
\begin{equation}
Q_{v'F',vF}^M = \langle \Psi_{v'F'M}
|\sum_i(\hat{f}_{iZ})^2| \Psi_{v FM} \rangle,  \label{e05_QvpFpvF}
\end{equation}
is crucial. One can prove that $Q_{v'F',vF}^M$ is nonzero only
if $v'-v=0$ or $\pm 2$ and $|F'-F|\leq 2$. Additionally, if
$M=0$, $(-1)^{F'}=(-1)^F$ is required. Of course, both the
pairs $(v',F')$ and $(v,F)$ should be symmetry-adapted. Let the
excitation caused by $Q_{v'F',vF}^M$ be called the elementary
excitation. For the ground state $\Psi_{v_g F_g 0}$, due to the
symmetry constraint\cite{r_D14}, one can prove that the
non-diagonal $Q_{v'F',v_g F_g}^0$ is nonzero only if
$(v',F')=(\bar{v}_g,\bar{F}_g)$, where $(\bar{v}_g,\bar{F}_g)
=(N,2N-2)$, $(2,2)$, and $(N-2,2)$, respectively, for the $f$-,
$p$-, and $c$-phases (refer to Tab.~\ref{tab1}). It implies
that each phase has only a unique mode of elementary
excitation. This is a distinguished feature of $M=0$ states.

When the ground state is not close to being degenerate, the
first-order perturbation theory can be used to calculate the
susceptibility. When $q'=0$ and $\varepsilon$ is very small,
due to having only a unique elementary mode, the first order
approximation of $\psi_{1,0}^{\varepsilon}$ can be written as
\begin{equation}
 \psi_{1,0}^{\varepsilon}
  =  \gamma
     ( \Psi_{v_g F_g,0}
      +\varepsilon
       \frac{Q_{\bar{v}_g\bar{F}_g,v_g F_g}^0}{E_{\mathrm{gap}}}
       \Psi_{\bar{v}_g\bar{F}_g,0}),
 \label{e06_psi10varepsilon}
\end{equation}
where $E_{\mathrm{gap}}=E_{\bar{v}_g\bar{F}_g}-E_{v_g F_g}$, $\gamma$ is
simply a constant for the normalization and can be easily obtained.
Inserting Eq.~(\ref{e06_psi10varepsilon}) into (\ref{e04_chiMqp}), we have
\begin{equation}
 \chi_0(0)
  \approx
     (\frac{Q_{\bar{v}_g\bar{F}_g,v_g F_g}^0}{E_{\mathrm{gap}}})^2.
 \label{e07_chi00}
\end{equation}

From Eq.~(\ref{e03_EvF}), we know that, for the $f$-phase
$E_{\mathrm{gap}}
=E_{N,2N-2}-E_{N,2N}=\frac{4N-1}{7}\sin\theta$, for the
$p$-phase
$E_{\mathrm{gap}}=E_{2,2}-E_{0,0}=\sin\theta-\cos\theta$, and
for the $c$-phase $E_{\mathrm{gap}}=E_{N-2,2}-E_{N,0}
=\frac{7\cos\theta-10\sin\theta}{70}(4N+2)-\frac{3\sin\theta}{7}$.
$E_{\mathrm{gap}}$ are also plotted in Fig.~\ref{fig1} by the
dash curves. In general, Eq.~(\ref{e07_chi00}) gives a very
good approximation except that $\theta$ is close to
$\theta_{pc}$ or $0\ (2\pi)$. When $\theta$ varies within a
region, $Q_{\bar{v}_g\bar{F}_g,v_g F_g}^0$ does not depend on
$\theta$. Therefore the $\theta$-dependence is due to the
factor $(1/E_{\mathrm{gap}})^2$ and can be shown by comparing
the solid and dash curves. Incidentally, since
$\Psi_{\bar{v}_g\bar{F}_g,0}$ might not be the first excited
state, the gap defined here might not be the energy difference
between the ground and the first excited states. Furthermore,
when $q'$ is not very weak, one has to go beyond the first
order perturbation theory, and therefore, in addition to the
elementary excitation, higher order excitations will be
included. Accordingly, various $\Psi_{v',F',0}$ components will
be contained in $\psi_{1,0}^{q'}$.

The features of the ground states with $M=0$ are summarized in
Tab.~\ref{tab1}. Some details in the table may be changed if
$N$ is not a multiple of 2 and 3 (Say, if $N$ is odd,
$\theta_{fp}$ is changed to
$\arctan[\frac{-7(N+4)}{10(N-1)}]<\pi$).

Note that the elementary excitation of the $f$-phase in region
I, namely, from ($v_g=N,F_g=2N$) to ($N,2N-2$), is essentially
a change of the total spin. The associated energy gap is nearly
proportional to $N$. Due to the enlargement of the gap by the
factor $N$, the $f$-phase has a very low susceptibility in
general. However, when $\theta\rightarrow 0$, the gap tends to
zero resulting in a great increase in the susceptibility. This
explains the sharp peak appearing in the left end of
Fig.~\ref{fig1}a or \ref{fig1}b. Nonetheless, since the gap
will increase with $N$, the sharp peak will be more and more
suppressed when $N$ is larger and larger.

The elementary excitation of the $p$-phase in region II,
namely, from ($0,0$) to (2,2), is a transformation of a singlet
pair into a $S=2$ pair. The associated $Q_{22,00}^0$ is roughly
proportional to $N$ \cite{r_D15}. Even when $N$ is not large,
it is still much larger than the $Q_{\bar{v}_g\bar{F}_g,v_g
F_g}^0$ of I and III as shown in Tab.~\ref{tab1}. Furthermore,
the gap is small and does not depend on $N$. Therefore, the
$p$-phase has a very high susceptibility, and will become
higher when $N$ is larger. The minimum of the gap appears at
the right border of II. Accordingly, there is a peak
\cite{r_B2011p}.

The elementary excitation of the $c$-phase in III, namely, from
($N,0$) to ($N-2,2$) is essentially a transformation of a
triplex to a singlet pair plus an extra particle. The
associated gap is also enlarged by $N$. It will be in general
large, therefore the $c$-phase has also a very low
susceptibility. However, the gap would decrease very fast if
$\theta$ is close to $\theta_{pc}$. Accordingly, there is a
sharp peak at the left border of III as shown in
Fig.~\ref{fig1}. Up to now the feature in Fig.~\ref{fig1} has
been explained.

Comparing Fig.~\ref{fig1}a with ~\ref{fig1}b we know that the
qualitative feature of the susceptibility does not change with
$N$. However, since $\theta_{fp}$ depends on $N$, the border
between I and II will move a little to the right when $N$
becomes larger and will tend to $\arctan(-7/10)
=145.01^{\circ}$ when $N\rightarrow \infty$ (this value
coincides with that from the MFT). Furthermore, the increase of
$N$ will cause a remarkable increase of $\chi_0(0)$ in II (due
to the fact that $Q_{2,2,0,0}^0$ is roughly proportional to
$N$) and will cause a small decrease of $\chi_0(0)$ in I and
III (Although $Q_{\bar{v}_g\bar{F}_g,v_g F_g}^0$ increases with
$N$, however the gap increase also. The two effects compete
with each other, and finally the latter over takes the former).

\section{Populations of spin components under a magnetic field}

When a magnetic field is applied, since the quadratic Zeeman term imposes
additional energy to the particles with $\mu\neq 0$, the ground state will
prefer to have the number of $\mu=0$ particles, $N_0$, larger so as to
reduce the energy. Thus, in addition to the susceptibility, the study of the
variation of $N_0$ is also a way to understand the effect of the field. Note
that the experimental measurement of the susceptibility might be difficult.
On the contrary, $N_0$ can be easily measured, say, via the Stern-Gerlach
technique.

Let the eigenstates of $H'_{\mathrm{mod}}$ be expanded in terms
of the Fock-states as $\psi_{iM}^{q'}
=\sum_{\alpha}c_{\alpha}^{q'iM}|\alpha\rangle$. One can extract
a particle (say, particle 1) from a Fock-state as
\begin{equation}
 |\alpha \rangle
  =  \sum_{\nu}
     \eta_{\nu}(1)
     \sqrt{\frac{N_{\nu}^{\alpha}}{N}}
     |\cdots,N_{\nu}^{\alpha}-1,\cdots \rangle,
 \label{e08_alpha}
\end{equation}
where $\eta_{\nu}$ is the spin-state of a particle in the
component $\nu$, $|\cdots ,N_{\nu}^{\alpha}-1,\cdots\rangle$ is
a Fock-state of the $(N-1)$-body system, in which the number of
particles in the component $\nu$ decreases by one. Therefore,
the particle can also be extracted from the eigenstate as
\begin{eqnarray}
 \psi_{iM}^{q'}
 &\equiv&
     \sum_{\nu}
     \eta_{\nu}(1)
     \psi_{\nu}^{q'iM},
 \label{e09_psiiMqp} \\
 \psi_{\nu}^{q'iM}
 &=& \sum_{\alpha}
     c_{\alpha}^{q'iM}
     \sqrt{\frac{N_{\nu}^{\alpha}}{N}}
     |\cdots,N_{\nu}^{\alpha}-1,\cdots\rangle.
 \label{e10_psivqpiM}
\end{eqnarray}
Thereby the probability of a particle in $\nu$ is just
\begin{equation}
 P_{\nu}^{i,M}
  \equiv
     \langle
     \psi_{\nu}^{q'iM}|
     \psi_{\nu}^{q'iM}
     \rangle.
 \label{e11_PviM}
\end{equation}
They fulfill $\sum_{\nu}P_{\nu}^{i,M}=1$. $P_{\nu}^{i,M}$ is
called the 1-body probability, and $NP_{\nu}^{i,M}
\equiv\bar{N}_{\nu}$ is just the average population of the
$\nu$ component of the $i$-th state.

\begin{figure}[htbp]
 \centering
 \resizebox{0.95\columnwidth}{!}{\includegraphics{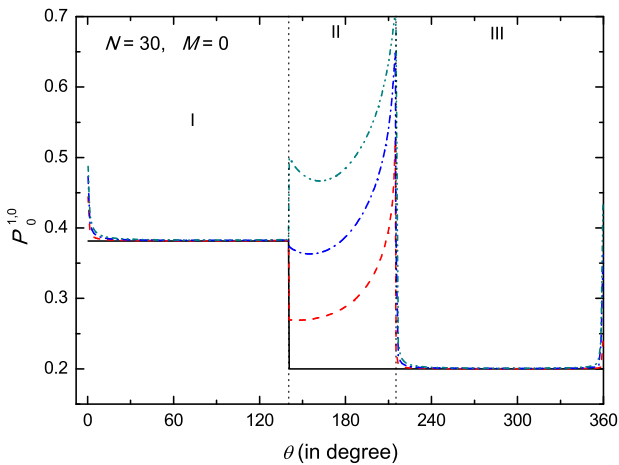}}
 \caption{(Color online.) $P_0^{1,0}$ against $\theta$ with
$N=30$ and $M=0$. The solid, dash, dash-dot and dash-dot-dot
curves have $q'=0$, $0.01$, $0.02$, and $0.03$, respectively.}
 \label{fig2}
\end{figure}

The one-body probabilities $P_0^{1,0}$ of the ground states
with $M=0$ and $q'$ given at a number of values are plotted in
Fig.~\ref{fig2}. When $q'=0$ (solid curve), the ground state
has three choices of phases as mentioned. The eigen-spin-state
of the $f$-phase is
\begin{equation}
\psi_{1,0}^{q'=0} = \Psi_{N,2N,0} = ( \{ [ (\eta\eta)_4 \eta ]_6
\eta \}_8 \cdots)_{2N,0}.  \label{e12_psi10qp0}
\end{equation}
where, the special way of spin coupling (i.e., the combined
spin of an arbitrary group of $j$ particles is $2j$) assures
that the spin-state written in Eq.~(\ref{e12_psi10qp0}) is
normalized and symmetrized. It is straight forward to extract a
particle from the form of Eq.~(\ref{e12_psi10qp0}), thereby the
1-body probability of the $f$-phase can be obtained as
$P_{\nu}^{1,0}=(C_{2,\nu;\ 2N-2,-\nu}^{2N,0})^2$, where the
Clebsch-Gordan coefficient has been introduced. When $\nu=0$,
$P_0^{1,0} =\frac{6N(2N-1)^2}{(4N-1)(4N-2)(4N-3)}$. This value
is equal to $0.381$ if $N=30$ as shown in Fig.~\ref{fig2}, and
will tend to $3/8=0.375$ if $N\rightarrow\infty$.

Both the $p$- and $c$-phases have $F=0$ (if $M=0$), and
therefore no special orientation is preferred. Accordingly, as
shown in \cite{r_HY2011}, $P_{\nu}^{1,0}=1/5$ for all $\nu$
disregarding how $N$ is. When the field is applied, both the
$f$- and $c$-phases are inert to the increase of $q'$, while
the $p$-phase is extremely sensitive. These coincide with the
previous findings in the susceptibility. In particular, when
$\theta$ is close to the right border of II, a great increase
of $\bar{N}_0$ appears even if the field is still weak. In I
and III the increase of $\bar{N}_0$ is very slight except when
$\theta$ is close to 0 ($2\pi$) and $\theta_{pc}$, where the
ground state is nearly degenerate as mentioned. The great
difference in sensitivity with respect to $q'$ holds
disregarding how $N$ is. Since the interactions of some
realistic spin-2 atoms (say, $^{87}$Rb, $^{23}$Na, and
$^{85}$Rb \cite{r_CCV2000}) might have
$\theta\approx\theta_{pc}$, the high sensitivity in this
particular domain would be very helpful to the exploration of
the interactions.

Incidentally, if $N$ increases from 30 to $\infty$, the general
feature of Fig.~\ref{fig2} would remain unchanged. However, the
border between the regions I and II would shift from
$140.48^{\circ}$ to $145.71^{\circ}$, the solid line in I (for
$q'=0$) would be lower a little from $0.381$ to $0.375$, the
parabolic-like curves in II for the $p$-phase would lie
remarkably higher (say, for $M=0$, $\theta=\pi$, and $q'=0.03$,
$P_0^{1,0}$ would be $0.282$, $0.485$, and $0.651$ if $N=6$,
$30$, and $60$, respectively).

\begin{figure}[htbp]
 \centering
 \resizebox{0.95\columnwidth}{!}{\includegraphics{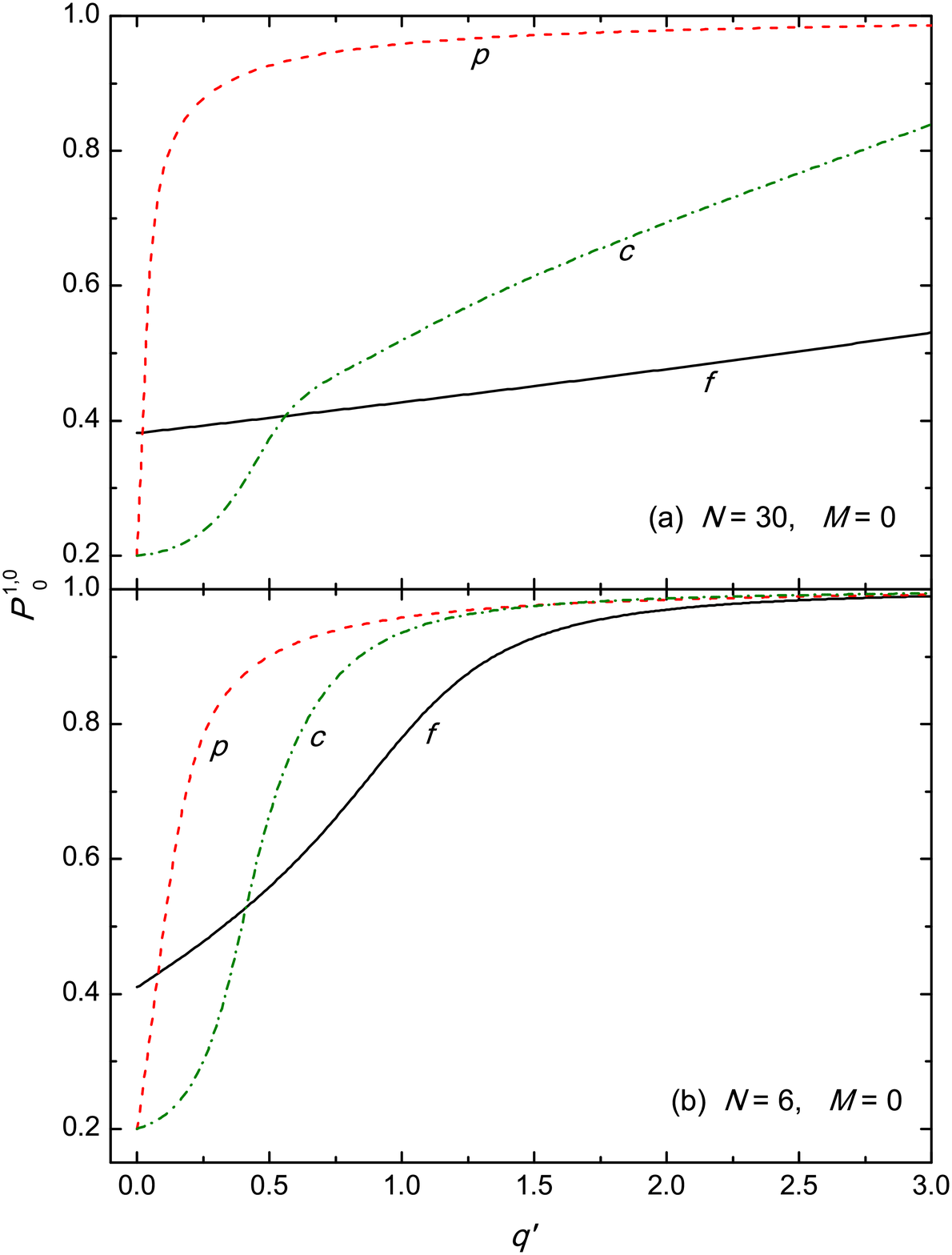}}
 \caption{(Color online.) $P_0^{1,0}$ against $q'$ with $N=30$
(a) and $6$ (b), and $M=0$. The solid, dash, and dash-dot
curves have $\theta=\pi/2$, $\pi$, and $3\pi/2$, and represent
the ferromagnetic ($f$), polar ($p$), and cyclic ($c$) phases,
respectively.}
 \label{fig3}
\end{figure}

It was found that, when $q'\rightarrow\infty$,
$\lim\langle0,0,N,0,0|\psi_{10}^{q'}\rangle^2=1$ disregarding
how $\theta$ is. It implies that all the ground states will
tend to the same Fock-state $|0,0,N,0,0\rangle$ as the strong
field limit, where all the spins have $\mu=0$. Accordingly,
$P_{\mu}^{1,0}\rightarrow 1$ (if $\mu=0$) or $\rightarrow 0$
(if $\mu\neq 0$). The process of going to the limit is shown in
Fig.~\ref{fig3}. Where $\theta$ is given at three values
$\pi/2$, $\pi$, and $3\pi/2$ to represent the $f$-, $p$-, and
$c$-phases, respectively. One can see that the dash curve for
the $p$-phase has a very steep take off at the left end. It
implies that the $p$-phase needs only a relatively much weaker
field to push it to its limit. This coincides with the finding
from Fig.~\ref{fig2}. When $N$ is larger, the take off is
steeper (e.g., when $q'$ increases from 0 to $0.25$,
$P_0^{1,0}$ would increase from $0.200$ to $0.783$, $0.877$,
and $0.905$, respectively, if $N=6$, 30, and 60). It implies
that, the $p$-phase of a larger condensate would be more
sensitive to the field as found before.

On the other hand, the take off of the dash-dot curve
($c$-phase) is much milder, and the solid curve ($f$-phase) is
the mildest, and they would be even milder if $N$ is larger.
For examples, when $q'=0.25$, the dash-dot curves would be
$0.300$, $0.237$, and $0.230$, and the solid curves would be
$0.478$, $0.393$, and $0.384$, if $N=6$, 30, and 60,
respectively. It implies that, the $f$- and $c$-phases of a
larger condensate would be more inert to the field than a
smaller one.

\section{The case with nonzero magnetization}

In the previous sections essentially the case with $M=0$ is
considered. When $q'=0$ and $M>0$, based on Eq.~(\ref{e03_EvF})
together with the constraints $2v\geq F\geq M$ and
$(-1)^N=(-1)^v$, the dependence of the quantum numbers of the
ground state $(v_g,F_g)$ on $\theta$ are listed in
Tab.~\ref{tab2}. The effect of $M$ is embodied by an integer
$v_0$, which is the smallest even (odd) integer $\geq M/2$ if
$N$ is even (odd), as shown in the table. Instead of three,
there are four phases. However, when $M=4k$ ($k=0,1,2,\cdots$)
and $N$ is even, or $M=4k+2$ and $N$ is odd, we have $2v_0=M $.
In this case, the two phases polar-A and -B are combined, and
the four phases reduce to three as before.

\begin{table*}[htbp]
\caption{The four phases of the ground states with $M\neq 0$.
Accordingly, the scope of $\theta$ is divided into four
connected regions, the right border of each region is given in
the last row. $v_0$ is the smallest even (odd) integer $\geq
M/2$ if $N$ is even (odd), and $(v_g,F_g)$ are the quantum
numbers of the ground states.}
\begin{ruledtabular}
  \label{tab2}
  \begin{tabular}{lllll}
  region       & I ($f$-phase)                                                        & II$_{\mathrm{A}}$ ($p\mathrm{A}$-phase) & II$_{\mathrm{B}}$ ($pB$-phase) & III ($c$-phase) \\
  \hline
  $(v_g,F_g)$  & $(N,2N)$                                                             & $(v_0,2v_0)$                            & $(v_0,M)$                      & $(N,M)$ \\
  right border & $\arctan\{\frac{-7[N(N+3)-v_0(v_0+3)]}{10[N(N-2)-v_0(v_0-2)]}\}<\pi$ & $\pi $                                  & $\arctan(\frac{7}{10})>\pi$    & $2\pi$
 \end{tabular}
 \end{ruledtabular}
\end{table*}

Note that, when $M$ is $\leq 5$, $(v_0,M)$ and $(N,M)$ might be
prohibited by symmetry \cite{r_D14}. Once $(v_0,M)$ or $(N,M)$
is prohibited, other more advantageous symmetry-adapted $(v,F)$
pair will replace it (Say, in II$_{\mathrm{B}}$ where a smaller
$v$ and a smaller $F$ will lead to a lower energy, if $(v_0,M)$
is prohibited, it would be replaced by $(v_0,M+1)$ and/or
$(v_0+2,M)$. In III where a larger $v$ and a smaller $F$ will
lead to a lower energy, if $(N,M)$ is prohibited, it would be
replaced by $(N-2,M) $ and/or $(N,M+1)$). In these cases the
appearance of more than four phases is possible.

Just for an example, we choose $N=$even and $M=7$ (this choice
is rather arbitrary). Then, we have $v_0=4$. Note that, for the
region II$_{\mathrm{B}}$, $(v_g,F_g)=(4,7)$ is prohibited.
Therefore, it would be replaced by $(4,8)$ and/or $(6,7)$. Note
that, from Eq.~(\ref{e03_EvF}), we would have $E_{4,8}=E_{6,7}$
if $\theta=\arctan(\frac{91}{170})=208.16^{\circ}$.
Accordingly, the II$_{\mathrm{B}}$ splits into two. When
$\pi\leq\theta <\arctan (\frac{91}{170})$, the symmetry-adapted
ground state will have $v_g=4,F_g=8$. Thus this part becomes an
extension of II$_{\mathrm{A}}$, and therefore the right border
of II$_{\mathrm{A}}$ extends from $\pi$ to
$\arctan(\frac{91}{170})$. When
$\arctan(\frac{91}{170})<\theta\leq\arctan(\frac{7}{10})=214.99^{\circ}$,
$v_g=6,F_g=7$. Thus the region II$_{\mathrm{B}}$ with the
quantum numbers $(6,7)$ becomes very narrow.

\begin{figure}[htbp]
 \centering
 \resizebox{0.95\columnwidth}{!}{\includegraphics{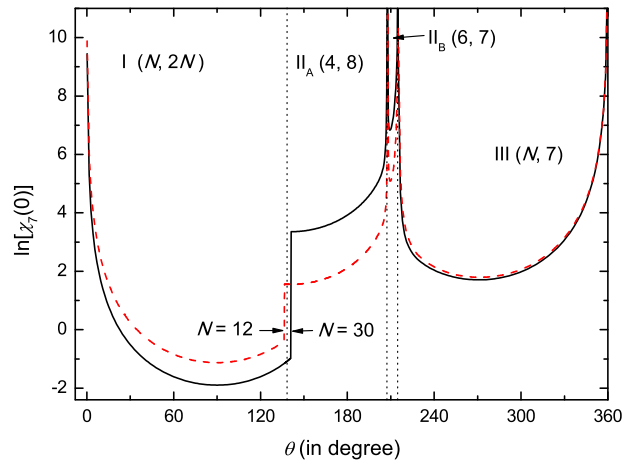}}
 \caption{(Color online.) $\ln(\chi_7(0))$ against $\theta$ with
$N=30$ (solid) and 12 (dash). $M=7$ is assumed. The domain of
$\theta$ are divided into four regions separated by the dotted
vertical lines (The dotted line at the left is simply the
average of the two cases with $N=30$ and 12). The quantum
numbers of the ground states $(v_g,F_g)$ in each region are
marked.}
 \label{fig4}
\end{figure}

The fidelity susceptibility against $\theta$ is shown in
Fig.~\ref{fig4}, where the division into four regions is clear
(including the very narrow region II$_{\mathrm{B}}$). In region
I for the $f$-phase, the eigenstate
$\Psi_{N,2N,M}=(\{[(\eta\eta)_4\eta]_6\eta\}_8 \cdots)_{2N,M}$,
where the spins are nearly lying along a common direction to
form a total spin, which has an arbitrary azimuthal angle but
not lying in the $X$-$Y$ plane. In II$_{\mathrm{A}}$ and
II$_{\mathrm{B}}$, the ground states are still dominated by the
singlet pairs together with $v_0$ unpaired particles. In
II$_{\mathrm{A}}$, the subsystem of the unpaired particles is
in the $f$-phase, i.e., they form a spin-state as
$(\{[(\eta\eta)_4\eta]_6\eta\}_8\cdots)_{2v_0,M}$. In
II$_{\mathrm{B}}$, the spins of the unpaired particles are
slightly diffused from a common direction, because their total
spin $M$ is slightly smaller than $2v_0$. In III, the $c$-phase
is dominated by the triplexes together with a few unpaired
particles. One can see that the susceptibilities of the
$p\mathrm{A}$- and $p\mathrm{B}$-phases are in general much
higher, and would become even higher if $N$ is larger. This is
similar to the findings from Fig.~\ref{fig1}. In particular,
when $\theta$ is close to the borders of the regions (except
the border between I and II$_{\mathrm{A}}$), the susceptibility
is exceptionally high.

\begin{figure}[htbp]
 \centering
 \resizebox{0.95\columnwidth}{!}{\includegraphics{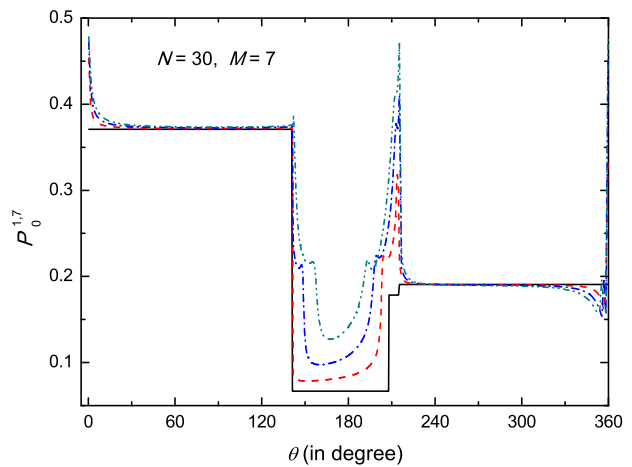}}
 \caption{(Color online.) $P_0^{1,7}$ against $\theta$ with
$M=7$ and $N=30$. The solid, dash, dash-dot and dash-dot-dot
curves have $q'=0$, $0.02$, $0.04$, and $0.06$, respectively.}
 \label{fig5}
\end{figure}

When $q'$ is given at four values the populations of $\mu=0$
components against $\theta$ are shown in Fig.~\ref{fig5}.
Similar to Fig.~\ref{fig2}, the $p\mathrm{A}$- and
$p\mathrm{B}$- phases respond to the increase of $q'$ more
sensitively. The set of curves in II$_{\mathrm{A}}$ and
II$_{\mathrm{B}}$ will become even higher when $N$ is larger.
In particular, when $\theta$ is close to 0 ($2\pi$), and close
to the narrow domain II$_{\mathrm{B}}$, the sensitivity is very
high. Obviously, the sensitivity emerging in Fig.~\ref{fig5} is
closely related to the susceptibility shown in Fig.~\ref{fig4}.

It was found that the number of nonzero non-diagonal matrix
elements $Q_{\bar{v}_g\bar{F}_g,v_g F_g}^M$ would be in general
more than one when $M\neq 0$. Therefore, there are more than
one elementary modes of excitation (i.e.,
Eq.~(\ref{e06_psi10varepsilon}) will contain more than two
terms). The actual mode is a mixing of them. It was found that,
when $q'$ is very small, the ground state $\psi_{1,M}^{q'}$ in
region I ($f$-phase), similar to
Eq.~(\ref{e06_psi10varepsilon}), is essentially
$\Psi_{N,2N,M}+b\Psi_{N,2N-2,M}$ where $b$ is a small constant.
For a numerical example, when $N=30$, $M=7$, $q'=0.005$ and
$\theta =\pi/2$, $b=-0.0019$. This value will become larger
when $\theta$ is close to zero (say, $b=0.1072$ if $\theta
=\pi/180$). In general, $|b|$ will decrease a little when $N$
becomes larger. In the region II$_{\mathrm{A}}$
($p\mathrm{A}$-phase) and for $M=7$ as an example,
$\psi_{1,7}^{q'}\approx \Psi_{4,8,7}+\sum_F b_F\Psi_{6,F,7}$,
where $F$ is ranged from 7 to 10 and $b_F$ are in the order of
1/10 or smaller. When $\theta $ is close to the left border of
II$_{\mathrm{A}}$, only $|b_{10}|$ is relatively larger
(meanwhile $\sin\theta$ is positive and the components with a
smaller $F$ will have a higher energy and therefore can be
neglected, refer to Eq.~(\ref{e03_EvF})). On the contrary, when
$\theta$ is close to the right border, only $|b_7|$ is
relatively larger (meanwhile $\sin\theta$ is negative). In
II$_{\mathrm{B}}$ ($p\mathrm{B}$-phase) $\psi_{1,7}^{q'}
\approx\Psi_{6,7,7}+b_{\alpha}\Psi_{6,8,7}+b_{\beta}\Psi_{8,7,7}$.
In III ($c$-phase) $\psi_{1,7}^{q'}
\approx\Psi_{N,7,7}+b_{\gamma}\Psi_{N-2,7,7}+b_{\delta}\Psi_{N,8,7}$.
The related coefficients in the two expansion are all small.
When $\theta$ is close to the left border of III, the gap
associated with the change of $v$ from $N$ to $N-2$ is very
small, and therefore $|b_{\gamma}|>>|b_{\delta}|$. On the
contrary, when $\theta$ is close to the right border $2\pi$,
the gap associated with the change of $F$ from $7$ to $8$ is
very small, therefore $|b_{\gamma}|<<|b_{\delta}|$.

\begin{figure}[htbp]
 \centering
 \resizebox{0.95\columnwidth}{!}{\includegraphics{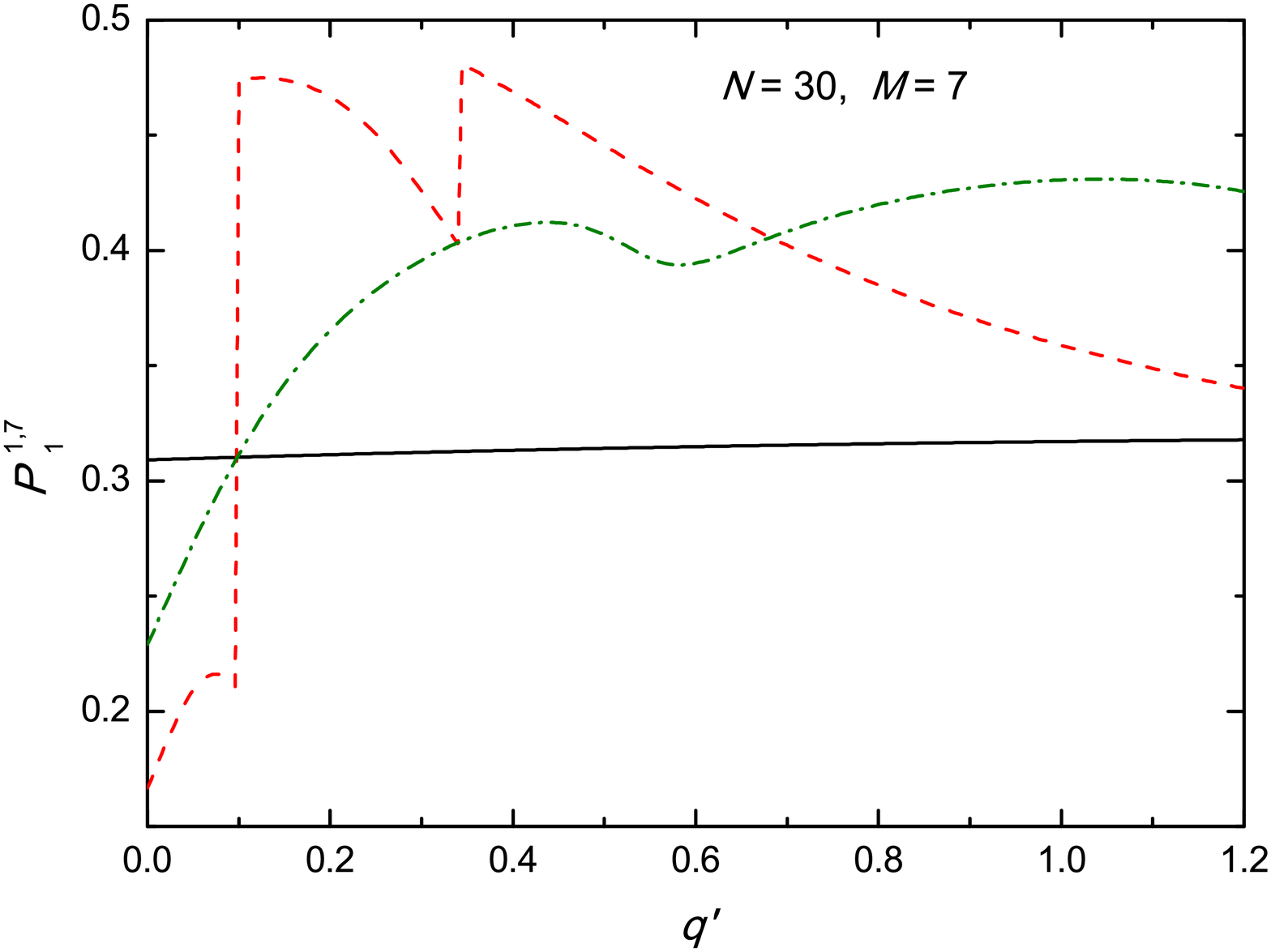}}
 \caption{(Color online.) $P_1^{1,7}$ against $q'$ with $M=7$
and $N=30$. The solid, dash, and dash-dot curves have
$\theta=\pi/3$, $\pi$, and $5\pi/3$ to represent the $f$-,
$p\mathrm{A}$-, and $c$-phases, respectively.}
 \label{fig6}
\end{figure}

\begin{figure}[htbp]
 \centering
 \resizebox{0.95\columnwidth}{!}{\includegraphics{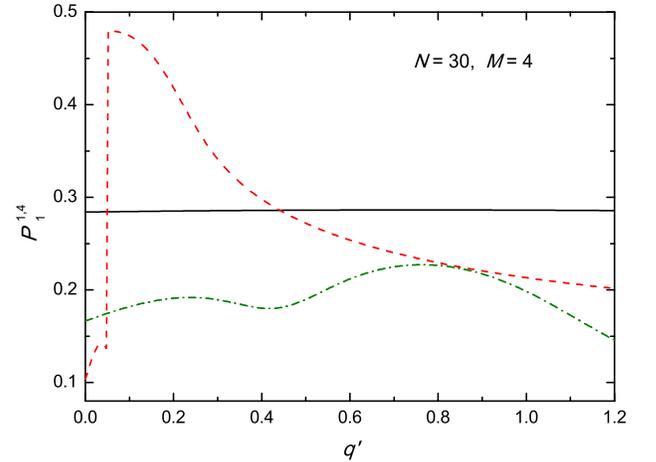}}
 \caption{(Color online.) $P_1^{1,4}$ against $q'$ with $M=4$
and $N=30$. The implications of the curves are the same as in
Fig.~\ref{fig6}. They are chosen to represent the $f$-, $p$-,
and $c$-phases.}
 \label{fig7}
\end{figure}

When $q'$ is large, it was found that
$\lim_{q'\rightarrow\infty}\langle
0,M,N-M,0,0|\psi_{1M}^{q'}\rangle^2=1$ disregarding how
$\theta$ is, i.e., all the phases tend to the same Fock-state
$|0,M,N-M,0,0\rangle$. Accordingly, $P_{\mu}^{1,M}\rightarrow
M/N$ (if $\mu=1$), $(N-M)/N$ (if $\mu=0$), and 0 (otherwise).
As examples, $P_1^{1,7}$and $P_1^{1,4}$ against $q'$ are shown
in Fig.~\ref{fig6} and Fig.~\ref{fig7}, respectively. In both
figures the solid curves ( $f$-phase) are inert to $q'$, the
dash-dot curves ($c$-phase) respond to $q'$ mildly. However,
the dash curves ($p\mathrm{A}$-phase in Fig.~\ref{fig6} and
$p$-phase in Fig.~\ref{fig7}) are more sensitive to $q'$. In
particular, at some critical points, abrupt changes are found
in the dash curves implying a sudden transition in
spin-structure (this abrupt change is not found in the previous
cases of $M=0$). In order to understand the transition better,
the associated ground state is expanded as
$\psi_{1,M}^{q'}=\sum_{v,F}b_{vF}\Psi_{vFM}$. Then, we define
$W_v=\sum_F(b_{vF})^2$ which is the weight of having $v$
unpaired particles. For the dash curve in Fig.~\ref{fig6}, we
found that $W_4=1$ at $q'=0$. When $q'$ increases, $W_4$
decreases and $W_6$ increases gradually. When $q'\rightarrow
q_{C_1}'=0.098$ (the critical point of the first transition),
$W_4\approx 0.55$ and $W_6\approx 0.39$. However, when $q'$
crosses the critical point, $W_4$ sudden becomes nearly zero
while $W_6$ jumps up to $0.90$. Thus the transition is
characterized by a great decrease of $W_4$ together with a
great increase of $W_6$. In Fig.~\ref{fig7}, the dash curve
have $W_2=1$ at $q'=0$. The first and unique critical point
$q'_{C_1}=0.050$. Similarly, the associated transition is
characterized by a great decrease of $W_2$ together with a
great increase of $W_4$. When $q'$ is larger than the range of
Figs.~\ref{fig6} and \ref{fig7}, no further critical points are
found. Instead, all the curves tend smoothly to their limit
$M/N$.

\section{Final remarks}

The effect of a magnetic field on the spin-structures of the ground states
of small spin-2 condensates is studied. The elementary modes of excitation
caused by the quadratic Zeeman term are found. The fidelity susceptibility
and the populations of spin-components are calculated. Mostly the case $N=30$
is studied in detail. The effect caused by increasing and decreasing $N$ is
also discussed. The following points are mentioned.

(i) The set of eigenstates $\Psi_{vFM}$ with the $U(5)\supset
SO(5)\supset SO(3)$ symmetry are introduced to help the
analysis. When the magnetic field is zero ($q'=0$) and $M=0$,
the pair of quantum numbers of the ground states $(v_g,F_g)$
have three choices associated with the three phases ($f$-,
$p$-, and $c$-phases).\cite{r_IPV2007} It turns out that, for
each phase, among all the non-diagonal matrix elements of the
quadratic Zeeman terms, only the one $Q_{\bar{v}_g\bar{F}_g,v_g
F_g}^0$ is nonzero. It implies each phase has its unique mode
of elementary excitation. Thus the response of the system to a
weak field is clear. In particular, the variation of the
fidelity susceptibility against $\theta$ can be understood in
an analytical way.

(ii) For the case $M\neq 0$, the ground state in general has four phases,
namely, the $f$-, $p\mathrm{A}$-, $p\mathrm{B}$-, and $c$-phases, each has
in general more than one elementary excitation modes. However, when $\theta$
locates at some particular domains, one elementary mode will be dominant.

(iii) The calculation of the fidelity susceptibility against
$\theta$ confirms the existence of the phases. The $p$-,
$p\mathrm{A}$- and $p\mathrm{B}$-phases have in general a much
higher susceptibility. In particular, when $\theta$ is close to
the borders separating the phases, the susceptibility may be
very high.

(iv) In addition to the fidelity susceptibility, the variation of the
populations against the field has also been calculated. Both quantities
provide similar message. In particular, the high sensitivity in the
neighborhood of the borders is found again in the population. Since it has
been suggested that a number of realistic species would have their
parameters of interaction close to the borders, \cite{r_CCV2000} since the
population is easier to measure, the high sensitivity is useful to the
accurate determination of the parameters of interactions.

(v) when $q'\rightarrow\infty$, all the ground states tend to
the limit $|0,M,N-M,0,0\rangle$ disregarding how $\theta$ is.
The process of tending to the limit is in general smooth.
However, when $M\neq 0$ and $q'$ is not very large (refer to
Fig.~\ref{fig6}), abrupt changes against $q'$ are found. It
implies that the field will cause not only a smooth variation
of structure but also a phase transition at some critical
points.

\begin{acknowledgments}
The suggestion on the calculation of the fidelity susceptibility by Prof.
Daoxin Yao is very much appreciated. The support from the NSFC under the
grant 10874249 is also appreciated.
\end{acknowledgments}


\begin{thebibliography}{99}
\bibitem{r_D14}         % 15
 Some essential points of symmetry constraint on the pair $(v,F)$ are the
following. (i) $F>2v$ is not allowed. (ii) When $N$ is even, $v$ must be
even. If $v$ is a multiple of 6 (including $v=0$), $F=1,2$, and 5 are not
allowed. If $v$ is not, $F=0,1$, and 3 are not allowed. (iii) When $N$ is
odd, $v$ must be odd. If $v-3$ is a multiple of 6, $F=1,2$, and 5 are not
allowed. If $v-3$ is not, $F=0,1$, and 3 are not allowed. (iv) $(v,2v-1)$ is
not allowed \cite{r_GA2004}.
\bibitem{r_D15}         % 16
 In general, the form of the eigenstate $\Psi_{vFM}$ is
complicated. However, for the $p$-phase, $\Psi_{000}$ is
exactly proportional to $\tilde{S}[(\eta\eta)_0]^{N/2}$, where
an even $N$ is assumed, $\tilde{S}$ is the operator of
symmetrization, and the spins are two-by-two coupled to zero.
$\Psi_{220}$ is exactly proportional to
$\tilde{S}(\eta\eta)_{20}[(\eta\eta)_0]^{N/2-1}$. Since they
have a rather simple form, they can be easily expanded in terms
of the Fock-states. Thereby the matrix element for the
elementary excitation, $Q_{22,00}^0$, can have an analytical
from. So, we can obtain numerical result even when $N$ is
larger (say, $N=500$), and we find that $Q_{22,00}^0$ is
roughly proportional to $N$.
\bibitem{r_B2011p}      % 17
 In fact, when $\theta$ is close to $\theta_{pc}$, not only the gap
decreases, a group of states having different $v$ but the same $F=F_g$ fall
down and the ground state becomes to be nearly degenerate. This is also an
important factor to raise the susceptibility.
\end{thebibliography}
\end{document}